# Real-Time Detection and Classification of Internal Leakage in Hydraulic Cylinders Using LSTM

1st Mehrbod Zarifi
Dept. of Mechanical Engineering
Amirkabir University of Technology
Tehran, Iran
M.Zarifi79@aut.ac.ir

2nd Mohamad Amin Jamshidi
Dept. of Mechanical Engineering
Amirkabir University of Technology
Tehran, Iran
m-a.jamshidi@aut.ac.ir

3rd Zolfa Anvari
New Technologies Research Center
Amirkabir University of Technology
Tehran, Iran
z.anvari@aut.ac.ir

4th Hamed Ghafarirad
Dept. of Mechanical Engineering
Amirkabir University of Technology
Tehran, Iran
Ghafarirad@aut.ac.ir

5th Mohammad Zareinejad
New Technologies Research Center
Amirkabir University of Technology
Tehran, Iran
mzare@aut.ac.ir

*Abstract*— **Hydraulic systems have been one of the most used technologies in many industries due to their reliance on incompressible fluids that facilitate energy and power transfer. Within such systems, hydraulic cylinders are prime devices that convert hydraulic energy into mechanical energy. One of the genuine and common problems related to hydraulic cylinders is leakages. Leakage in hydraulic systems can cause a drop in pressure, general inefficiency, and even complete failure of such systems. The various ways leakage can occur define the major categorization of leakage: internal and external leakage. External leakage is easily noticeable, while internal leakage, which involves fluid movement between pressure chambers, can be harder to detect and may gradually impact system performance without obvious signs. When leakage surpasses acceptable limits, it is classified as a fault or failure. In such cases, leakage is divided into three categories: no leakage, low leakage, and high leakage. The main idea of this work is a defect detection algorithm whose primary duty is to find the least amount of leakage in the hydraulic system at the shortest possible detection time. To fully develop this idea, experimental data collection of Hydraulic systems is required. The collected data uses pressure sensors and other signals that are single-related. Due to the utilization of Long Short-Term Memory (LSTM) recurrent neural networks, more complex data analysis was enabled, which the LSTM-based leakage detection algorithm successfully achieved, providing 96% accuracy in classifying leakage types. The results show that the suggested approach can diagnose faults in real-time with an average of 5ms per cycle. Early failure detection lowers maintenance costs and increases the longevity of the hydraulic system.**

*Keywords*— **Hydraulic Systems, Internal Cylinder Leakage, Experimental Data Collection, Fault Detection, Recurrent Neural Networks.**

## I. Introduction

Hydraulic systems have long been recognized as one of the most important and widely used technologies across various industries, operating based on the use of incompressible fluids for the transfer of power and energy. The unique features of these systems, including their ability to generate very large forces and precisely control movement, have made them especially useful in heavy and sensitive industries. These capabilities are attributed to the incompressibility of fluids, meaning that even small changes in the fluid volume can result in significant pressure variations, which are then converted into mechanical motion.

Hydraulic cylinders, as fundamental components in hydraulic systems, are responsible for converting hydraulic energy into mechanical energy. These critical parts, used in many industrial machines and heavy equipment, generate immense forces in a specific direction through linear motion. The proper functioning of these cylinders is crucial for maintaining the overall efficiency and performance of hydraulic systems.

Leakage in hydraulic cylinders can be categorized into two types: internal and external, depending on their location and source. Internal leakage occurs when the fluid within the cylinder flows from a high-pressure chamber to a low-pressure chamber without escaping the system. This type of leak is often difficult to detect and can gradually lead to decreased system performance over time. In contrast, external leakage happens when hydraulic fluid seeps out of the system and into the external environment. These leaks are easily noticeable, typically manifesting as drops of fluid around the joints or fittings of the cylinder.

Detecting leaks and their locations in hydraulic systems is a branch of control engineering that can be divided into two main approaches:

*A. Model-based Methods*

This approach employs mathematical models and simulations of the system's behavior to analyze unusual signals. These methods can accurately identify leaks, particularly internal leaks, which are often challenging to detect. However, they require precise mathematical models, and complete and accurate datasets, and are highly sensitive to modeling errors, which can limit their practical application in real-world scenarios [1]. Moreover, these methods face additional challenges, including the inability to be applied in closed-loop systems and reduced algorithm sensitivity due to the dispersion of effects on coefficients after multiple transformations.

In 2002, Hong-Zhou Tan and Sepehri [2] proposed a model-based fault diagnosis method for electrohydraulic systems using a second-truncated Volterra nonlinear model. This method detects and isolates system faults by analyzing input-output signals, offering efficient monitoring and fault detection for hydraulic systems. Despite its effectiveness, the reliance on a highly accurate nonlinear model makes it susceptible to inaccuracies caused by incomplete data or modeling assumptions.

Similarly, in 2005, Khan et al. [3] introduced a nonlinear observer-based fault detection technique for electro-hydraulic servo-positioning systems. By employing a nonlinear observer and Wald's sequential test, the method detects faults such as incorrect pump pressure and sensor errors in real time. While innovative, the approach demands precise model calibration and is sensitive to errors in the observer design, which can reduce its robustness in varied operational conditions.

In 2011, Mahulkar et al. [4] proposed a fault identification method for hydraulic systems using derivative-free filtering. This approach models faults as parametric errors and estimates them using a divided difference filter (DDF). The experimental setup simulated various faults, focusing on internal leakages. The method successfully identified faults as small as 0.01 l/s and demonstrated robustness to disturbances like friction. However, distinguishing between faults represented by additive parameters in the same equation remains challenging. Sensitivity analysis revealed that the process noise covariance significantly affects fault estimation accuracy.

In 2023, Saeedzadeh et al. [5] developed a model-based fault detection and diagnosis (FDD) strategy for electro-hydraulic actuators (EHAs) by integrating an updated interactive multiple model (UIMM) with a smooth variable structure filter with a variable boundary layer (SVSF-VBL). This approach reduces computational load by limiting the number of active models during fault detection and isolation, making it suitable for real-time applications. While effective in identifying leakage and friction faults, the method's reliance on precise models and sensitivity to uncertainties in dynamic systems highlight potential limitations in environments with incomplete data or significant system variability.

*B. Signal-based Methods*

These methods utilize sensors to measure and analyze the system's output signals. By comparing sensor data with reference and standard values, any deviations from normal operation are identified, allowing for the detection of potential leaks. However, these methods are sensitive to noise in the sensor data and require large amounts of data for accurate analysis, which can pose challenges in practical applications [6].

In 2010, Yazdanpanah Goharrizi and Sepehri [7] developed a wavelet-based method to detect internal leakage in hydraulic actuators by analyzing pressure signals. The method successfully identifies even small leaks without needing a model of the actuator. Also in 2011, Yazdanpanah Goharrizi et al. [8] introduced a wavelet-based approach for online diagnosis of external leakage in hydraulic actuators and its isolation from internal leakage. By analyzing pressure signals using wavelet coefficients, the method successfully detects small leakages during operation without requiring a model. And in 2012, Yazdanpanah Goharrizi and Sepehri [9] applied the Hilbert-Huang Transform (HHT) to detect internal leakage in hydraulic actuators. By analyzing pressure signals through empirical mode decomposition and Hilbert spectrum, the method successfully detected small leakages without needing a system model.

*C. AI-based Methods*

Artificial intelligence plays a crucial role in detecting hydraulic cylinder leaks. By utilizing machine learning algorithms and artificial neural networks, intelligent systems can analyze sensor data and identify abnormal patterns. These analyses assist in early leak detection and help prevent more serious failures. One of the main advantages of AI is its high accuracy in detecting small and internal leaks, which are difficult to identify with traditional methods. Moreover, AI is capable of analyzing large and complex datasets, enabling it to detect long-term trends that contribute to preventing potential breakdowns.

AI-based systems for leak detection offer numerous benefits. They not only enhance detection accuracy but also allow for continuous, real-time monitoring of systems. With this technology, the need for frequent stops for inspection and maintenance is reduced, leading to improved system efficiency. Furthermore, these systems help lower repair costs, extend the service life of hydraulic systems, and mitigate safety risks. They can also analyze data to provide optimized solutions to problems, ultimately improving overall system performance.

In 2016, Sharifi et al. [10] developed a signal-based fault detection method for electro-hydraulic servo systems using support vector machines (SVM). The approach analyzed pressure signals in response to periodic square inputs to identify internal leakage. Features such as the height, width, and location of pressure peaks were extracted for classification into multiple fault levels. This method demonstrated high classification accuracy while reducing computational costs through dimensionality reduction. However, it relies on feature extraction, making it sensitive to data variations and noise.

In 2018, Sharifi et al. [11] proposed a leakage fault detection method for electro-hydraulic servo systems using a nonlinear representation learning approach, extending their earlier work. The method extracts features such as the height, width, and location of pressure peaks and transforms these features into a new coordinate system using a two-layer neural network. By maximizing separability in the transformed space through a custom optimization algorithm, the method significantly reduces classification errors and computational costs. Experimental results demonstrated high accuracy in detecting internal leaks under various conditions, although the approach is sensitive to noise and requires substantial computational resources. Also in 2021, Jose et al. [12] proposed an early detection and classification method for internal leakage in the boom actuators of mobile hydraulic machines. The method uses pressure and boom angle signals, with features extracted and selected through Particle Swarm Optimization (PSO). The Support Vector Machine (SVM) classifier achieved 97% accuracy in detecting leakage levels. And in 2024, Takeda et al. [13] developed a machine-learning-based system for detecting abnormal pressure drops in hydraulic press machines. The system uses vibration sensors and Gaussian Process Regression (GPR) to model normal operations and detect deviations caused by failures, such as oil leakage.

The goal of this paper is to develop a real-time and online leakage detection system for each cycle using a recurrent neural network (LSTM) that accurately classifies hydraulic system conditions based on time-dependent pressure signals. By reducing computational complexity and emphasizing mechanical significance, the system can detect and classify leaks into no leakage, minor leakage, and major leakage categories, enabling early intervention to prevent further damage in industrial applications.

## II. EXPERIMENTAL SETUP

### A. The Schematic Diagram of A Hydraulic Actuator Internal Leakage

Fig. 1. presents the schematic diagram of a hydraulic actuator internal leakage. The parameter $q_1$ represents the internal leakage resulting from seal damage, which causes fluid to flow from the left chamber to the right chamber.

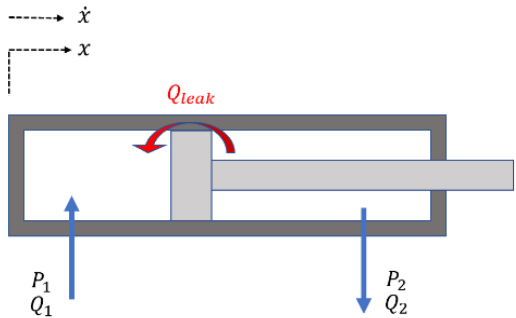

Fig. 1. The schematic diagram of the internal leakage

### B. The Schematic Diagram of The Experimental Hydraulic Setup

A schematic representation of the experimental setup is illustrated in Fig. 2. The system utilizes a double-acting cylinder as the actuator. The fluid pressure is supplied by a pumping unit.

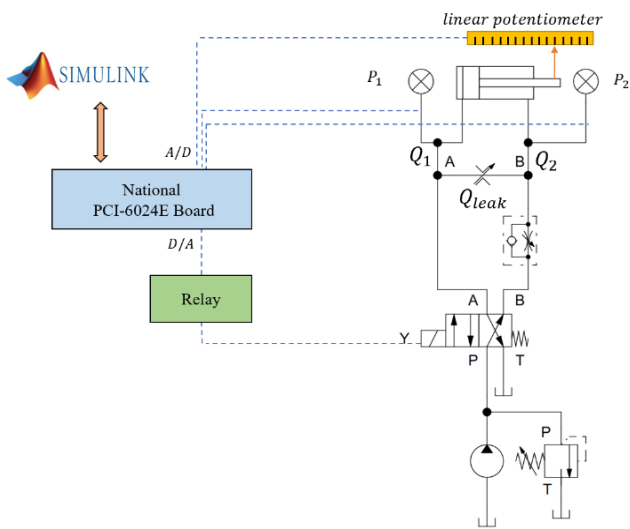

Fig. 2. The schematic diagram of the experimental hydraulic system

### C. The Experimental Hydraulic Setup

Fig. 3. illustrates the experimental setup, which was assembled in the Robotics and Servo Systems Laboratory at the New Technologies Research Center of Amirkabir University of Technology. This setup comprises a Bosch Hydraulic Cylinder H160CA 63×28, a FESTO Four-way two-position directional control valve, an A10VSO Series 31 Bosch Rexroth Axial Piston Variable Pump, a SIEMENS INNOMOTICS SD 1LE1 Standard Motor, a Bosch Rexroth Pressure Relief Valve DB 20-3-50, a FESTO pressure sensor, and a Trafag AG ECT 8472 Industrial Pressure Transmitter.

The piston position is measured using a RTL-450-D-5K-C1 OPKON Linear potentiometer attached to the hydraulic cylinder. As depicted in Fig. 3., leakage is artificially induced by connecting both chambers of the cylinder through an open flow control valve. The pressure signals from the left chamber ($P_1$) are utilized for leakage detection.

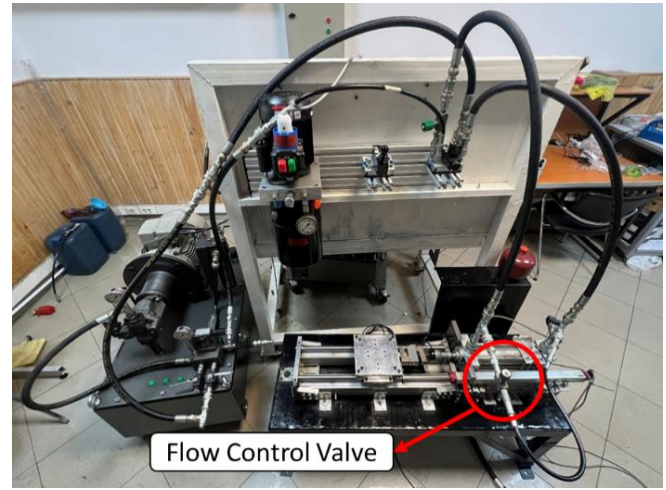

Fig. 3. The experimental hydraulic setup

### D. Performance Analysis of The Laboratory Hydraulic System

To analyze the differences in the performance of the hydraulic system, an experimental test was conducted where pressure signals were recorded after inducing artificial leakage using a flow control valve. The hydraulic cylinder was moved from the start to the end of its stroke under different commands, and this process was repeated for three conditions: no leakage, low leakage, and high leakage. As shown in Fig. 4., the pressure signals for these three conditions are displayed, having been recorded after inducing artificial leakage, allowing for clear comparison and analysis of the differences between the scenarios.

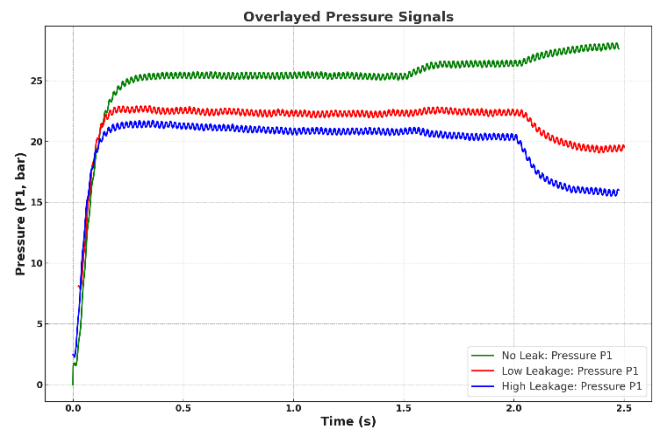

Fig. 4. Pressure Signals of Healthy Syestm, Low Fult, and High Fault Sytems

As can be seen, as the leakage increases, the pressure decreases accordingly. This happens because a regenerative circuit is created within the system. In Fig. 5., it can be seen what occurs within the regenerative circuit.

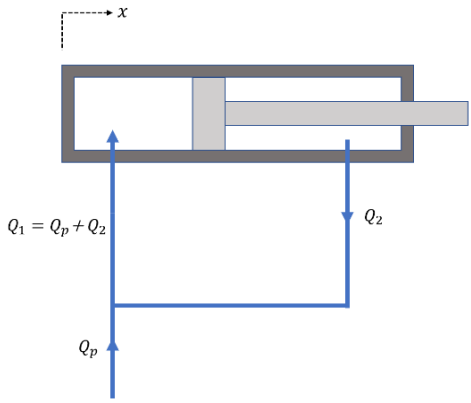

Fig. 5. Regenerative Circuit

As shown in Fig. 5., the flow rate behind the cylinder increases as leakage grows. Since the cylinder's speed is calculated using the equation $V = \frac{Q}{A}$, A higher leakage results in an increased flow rate behind the cylinder, which in turn leads to a faster cylinder speed. Consequently, the number of back-and-forth cycles during the stroke decreases. Additionally, the creation of a regenerative circuit, while increasing the cylinder's speed, causes a pressure drop, reducing the force the cylinder can exert. This reduction in force can negatively impact performance, especially in situations where high force is required. This difference is clearly shown in Fig. 8.

### III. Fault Detection Algorithm

In hydraulic systems, internal leakage reduces the pressure difference between the two chambers of the cylinder, leading to a damping effect on the system's dynamics. This alters the transient response of the pressure signals. Each time the direction of the hydraulic actuator changes, pressure peaks emerge, making these signals valuable for detecting internal leakage. In this research, experimental data collection was conducted to capture the time-series pressure signals from the rear chamber of the cylinder, with special attention to the peaks within these signals. A Long Short-Term Memory (LSTM) network was then employed for classification.

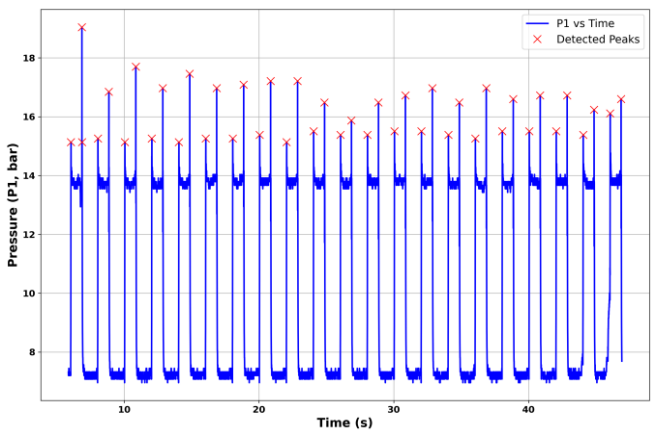

Fig. 6. Left Chamber Pressure Signal With Detected Peaks for Healthy System

The pressure signals, as illustrated in Fig. 6, along with their peaks, were processed directly as time series data. This enabled the LSTM network to learn both overall trends and critical peak information, leading to accurate leakage detection with minimal error. The proposed fault detection flowchart, as depicted in Fig. 7, outlines the structure of this approach, providing a reliable system for real-time internal leakage identification. This flowchart illustrates how a leakage detection algorithm using recurrent neural networks (LSTM) works. First, pressure signals are collected and pre-processed. Peak detection is then performed, and the data is split into training and validation sets. The neural network is trained using the training data, and its performance is validated with the validation data. Finally, the system is classified as either healthy or faulty. If the results are unsatisfactory, the neural network's hyperparameters are returned.

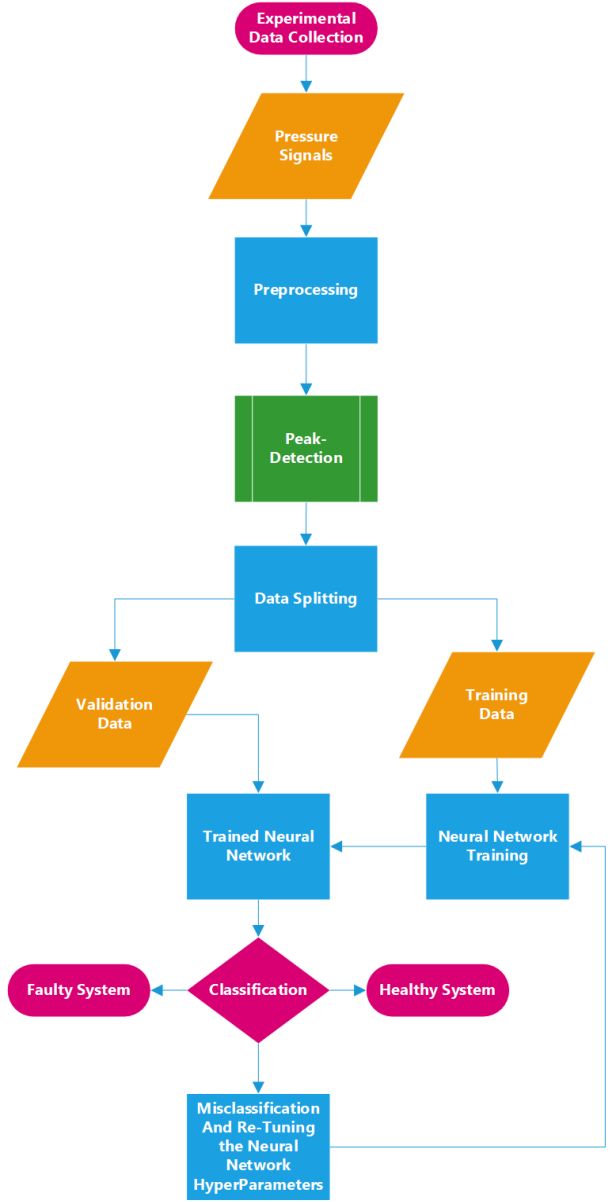

Fig. 7. The schematic diagram of the structure of the fault detection algorithm

#### A. LSTM Model

In this research, a Long Short-Term Memory (LSTM) network was employed for time-series data classification. LSTM, as a type of recurrent neural network, is designed to capture long-term dependencies between time-series data and to overcome the short-term memory issue typically found in standard recurrent neural networks.

The model consists of seven layers: LSTM, Dense, and Dropout, each serving a specific purpose. The LSTM layers are responsible for extracting temporal features, the Dense layers reduce complexity, and the Dropout layers help prevent overfitting. These layers and hyperparameters were determined after extensive trial and error, ensuring the architecture was optimized for achieving the best performance. Together, they work for the final classification of time-series data into three categories: no leakage, low leakage, and high leakage. For detailed information on the structure and purpose of each layer, please refer to Table 1 below.

TABLE I. OPTIMIZED LSTM MODEL

| Layer | Neuron Count | Activation Function | Description |
|---|---|---|---|
| First LSTM Layer | 128 | N/A (LSTM) | Extracts temporal features, passes outputs from each time step. |
| First Dropout Layer | N/A (30% dropout rate) | N/A | Prevents overfitting by deactivating neurons during training. |
| Second LSTM Layer | 64 | N/A (LSTM) | Focuses on final output, reduces complexity. |
| Second Dropout Layer | N/A (30% dropout rate) | N/A | Prevents overfitting by deactivating neurons during training. |
| First Dense Layer | 128 | ReLU | Processes learned features, connects them nonlinearly. |
| Second Dense Layer | 64 | ReLU | Reduces complexity, focuses on most relevant features. |
| Output Dense Layer | 3 | Softmax | Classifies data into 3 categories using softmax activation. |

After the LSTM layers, two fully connected Dense layers are used to further process the learned features. The first Dense layer contains 128 neurons, and the second reduces complexity with 64 neurons. Both use the ReLU activation function to capture nonlinear relationships in the data. The final output layer, a Dense layer with 3 neurons, uses softmax activation to classify the data into the three leakage categories based on probabilistic outputs.

To optimize the model, the Adam optimization algorithm is employed. This algorithm automatically adjusts the learning rate to ensure the model trains effectively. In this model, the learning rate is set to 0.0003. Additionally, the categorical cross-entropy loss function is used to calculate the model's error, specifically designed for multi-class classification problems. This loss function measures the difference between the model's predictions and the actual labels, helping to optimize the model accordingly. Finally, the model is evaluated using accuracy as a metric to determine how well it performs in making correct classifications.

### B. Results

According to Fig. 8, it can be seen that distinguishing between the high leakage data and the other two classes is significantly easier. In contrast, the data from the no leakage and low leakage classes show less differentiation, making it more challenging to separate them. Consequently, the model's accuracy for classifying these two classes will likely be lower than for the high leakage data.

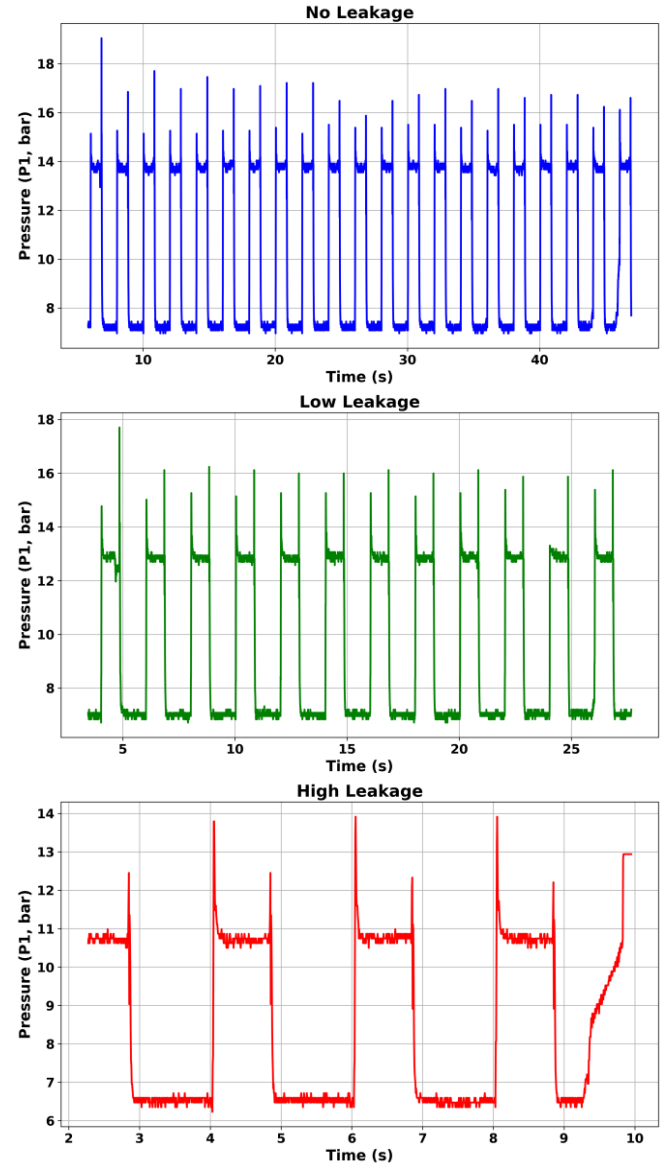

Fig. 8. Full pressure signals of a healthy system, Low leakage, and high leakage systems

Next, to validate these claims, tests were conducted, and the results are examined as follows:

The training process spanned 200 epochs, during which the model demonstrated notable improvements in both accuracy and loss metrics. 80% of the data was used for training, and 20% was used for testing. Initially, the training accuracy was 62.44%, with a loss of 0.7210. As the training progressed, the accuracy steadily increased, reaching approximately 96.06% by the final epoch, while the loss decreased to 0.0737. Validation accuracy also followed a positive trend, starting at 73.27% in the first epoch and peaking at 96.96% by the last epoch. Similarly, the validation loss dropped from 0.6248 to 0.0867 by the end of training.

The left chart in Fig. 9 reflects the accuracy improvement over the epochs, confirming the model's effectiveness in learning and classifying the data. The right chart illustrates the decline in training and validation loss, showing a well-optimized model. Additionally, the model achieved a test accuracy of 95.81%, indicating strong generalization to unseen data and confirming its capability to distinguish between different classes effectively. These results suggest that the model performed successfully in the intended classification task.

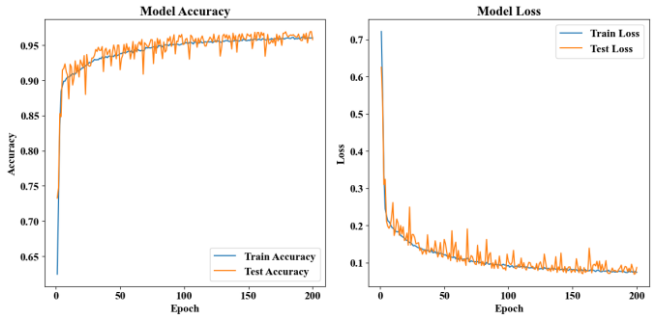

Fig. 9. Train and test

These results demonstrate the model's ability to accurately differentiate between the different classes, indicating strong performance in the classification task.

According to Fig. 10, the model's performance can be assessed across three classes: no leakage (0), low leakage (1), and high leakage (2).

- For the "no leakage" class (0), the model correctly predicted 3,890 instances, misclassifying 131 as low leakage and 0 as high leakage.

- In the "low leakage" class (1), the model accurately identified 6,768 instances but misclassified 362 as no leakage and 4 as high leakage.

- Lastly, for the "high leakage" class (2), the model correctly predicted 3,701 instances, while 2 were misclassified as low leakage.

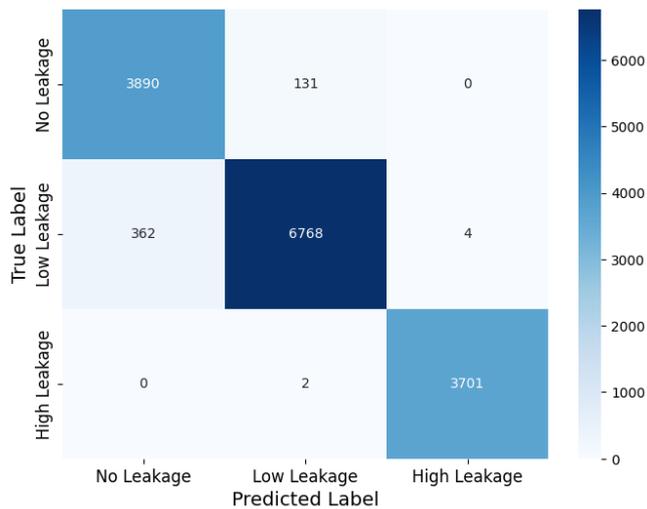

Fig. 10. Confussion Matrix

Overall, the model demonstrates strong classification performance, particularly for the "high leakage" and "no leakage" classes, with a significant number of correct predictions and relatively few misclassifications. The confusion matrix highlights areas where the model excels and where there might be some challenges, especially in distinguishing between the low leakage and no leakage classes.

According to Fig. 11, the precision-recall curve provides insights into the model's performance across three classes: no leakage (Class 0), low leakage (Class 1), and high leakage (Class 2).

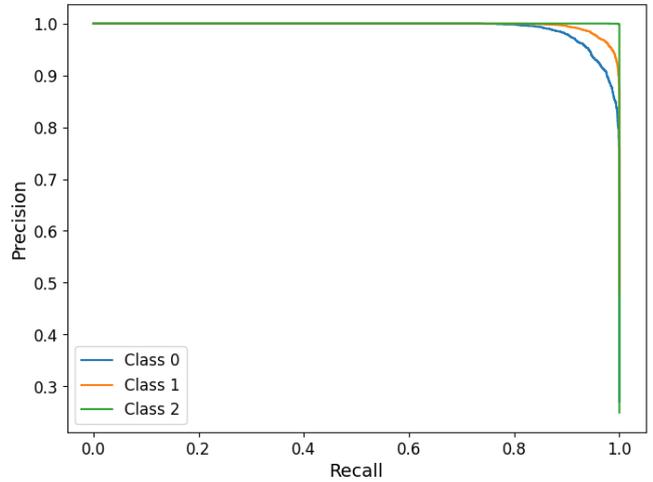

Fig. 11. Precision-Recall Curve

The curves for each class indicate that as the recall increases towards 1.0, the precision remains relatively high, particularly for Class 2 (high leakage), which suggests that the model is very effective at identifying instances of high leakage without generating many false positives.

In contrast, Classes 0 and 1 show a slight drop in precision as recall approaches 1.0, indicating some trade-offs between precision and recall, especially for low leakage. Overall, these results reflect the model's ability to maintain a good balance between precision and recall, demonstrating its robustness in classifying the different leakage scenarios.

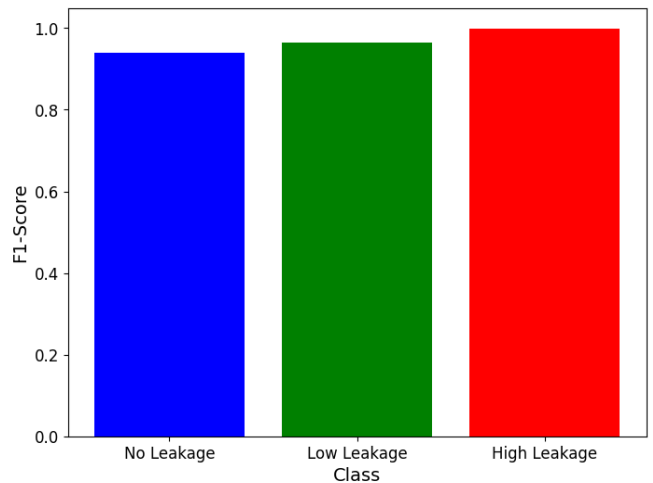

Fig. 12. F1-Score for each class

According to Fig. 12, the F1 scores for each class are depicted, highlighting the model's performance in classifying the different leakage scenarios. The F1 scores for the three classes are as follows:

- Class 0 (No Leakage): The F1-score is approximately 0.9404, indicating a strong balance between precision and recall in identifying instances without leakage.

- Class 1 (Low Leakage): The F1-score is around 0.964446, showing that the model is particularly effective at correctly classifying low leakage instances, achieving an even better balance between precision and recall.

- Class 2 (High Leakage): The F1-score is nearly 1.000 (0.999190), suggesting that the model excels in identifying high leakage instances with a high degree of accuracy and minimal false positives.

Overall, these results reflect the model's robust classification capability across all classes, with the highest performance observed for Class 2, demonstrating its effectiveness in detecting high leakage scenarios.

## IV. CONCLUSION

In summary, this research successfully developed and implemented an intelligent algorithm for detecting internal leaks in hydraulic systems, leveraging advanced deep-learning techniques to achieve high accuracy and reliability. The primary focus was on designing a robust fault detection method capable of real-time monitoring, precise leak classification, and low computational demands.

The study introduced an LSTM algorithm, optimized using the Optuna library, which achieved exceptional accuracy of 97.00% through hyperparameter tuning and 96.07% with cross-validation. This high accuracy underscores the algorithm's reliability and robustness, as even without optimization, it delivered comparable results. These results show how well the system detects and categorizes leaks, even small or imperceptible ones, guaranteeing real-time fault detection and leakage classification with a 5ms latency for each operational cycle. This capability is crucial for maintaining system performance, enabling timely responses, and reducing maintenance costs.

In addition to the advancements with RNNs, experimental data collection from a laboratory hydraulic system played a pivotal role in this research. By including pressure signals and sensor readings under various leakage conditions, the study ensured the algorithm's ability to learn complex patterns and distinguish between no leakage, low leakage, and high leakage scenarios. After training recurrent neural networks (LSTM) on these datasets, the algorithm achieved an impressive accuracy of 95.81% in identifying leaks, further validating the effectiveness of combining precise data collection with powerful deep learning tools.

This research highlights the importance of real-time leak detection in preventing hydraulic system failures, which can lead to reduced performance, energy waste, and catastrophic failures. By ensuring continuous monitoring and reliable fault classification, the proposed algorithm significantly enhances system efficiency and reliability across diverse industrial applications.

Overall, this work represents a significant step forward in leveraging artificial intelligence for hydraulic system maintenance. The proposed RNN-based fault detection algorithm demonstrates the potential of intelligent systems to revolutionize hydraulic system monitoring, providing timely and precise leak detection to prevent failures, reduce downtime, and optimize operational performance. Implementing this algorithm across various industries can greatly enhance the reliability and longevity of hydraulic systems.

REFERENCES

[1] S. Simani, C. Fantuzzi, R. J. Patton, S. Simani, C. Fantuzzi, and R. J. Patton, Model-based fault diagnosis techniques. Springer, 2003.

[2] H.-Z. Tan and N. Sepehri, "Parametric fault diagnosis for electrohydraulic cylinder drive units," IEEE Transactions on Industrial Electronics, vol. 49, no. 1, pp. 96-106, 2002.

[3] Khan, H., Abou, S. C., & Sepehri, N. (2005). Nonlinear observer-based fault detection technique for electro-hydraulic servo-positioning systems. Mechatronics, 15(9), 1037-1059.

[4] V. Mahulkar, D. E. Adams, and M. Derriso, "Derivative free filtering in hydraulic systems for fault identification," Control engineering practice, vol. 19, no. 7, pp. 649-657, 2011.

[5] A. Saeedzadeh, S. Habibi, M. Alavi, and P. Setoodeh, "A Robust Model-Based Strategy for Real-Time Fault Detection and Diagnosis in an Electro-Hydraulic Actuator Using Updated Interactive Multiple Model Smooth Variable Structure Filter," *Journal of Dynamic Systems, Measurement, and Control,* vol. 145, no. 10, 2023, doi: 10.1115/1.4063206.

[6] G. J. Vachtsevanos, F. Lewis, M. Roemer, A. Hess, and B. Wu, Intelligent fault diagnosis and prognosis for engineering systems. Wiley Online Library, 2006.

[7] A. Y. Goharrizi and N. Sepehri, "A wavelet-based approach to internal seal damage diagnosis in hydraulic actuators," IEEE transactions on industrial electronics, vol. 57, no. 5, pp. 1755-1763, 2009.

[8] A. Y. Goharrizi, N. Sepehri, and Y. Wu, "A wavelet-based approach for online external leakage diagnosis and isolation from internal leakage in hydraulic actuators," International journal of fluid power, vol. 12, no. 2, pp. 37-47, 2011.

[9] A. Y. Goharrizi and N. Sepehri, "Internal leakage detection in hydraulic actuators using empirical mode decomposition and Hilbert spectrum," IEEE Transactions on Instrumentation and Measurement, vol. 61, no. 2, pp. 368-378, 2011.

[10] S. Sharifi, S. M. Rezaei, A. Tivay, F. Soleymani, and M. Zareinejad, "Multi-class fault detection in electro-hydraulic servo systems using support vector machines," in 2016 4th International Conference on Robotics and Mechatronics (ICROM), 2016: IEEE, pp. 252-257.

[11] S. Sharifi, A. Tivay, S. M. Rezaei, M. Zareinejad, and B. Mollaei-Dariani, "Leakage fault detection in Electro-Hydraulic Servo Systems using a nonlinear representation learning approach," ISA transactions, vol. 73, pp. 154-164, 2018.

[12] J. T. Jose, J. Das, S. K. Mishra, and G. Wrat, "Early detection and classification of internal leakage in boom actuator of mobile hydraulic machines using SVM," Engineering Applications of Artificial Intelligence, vol. 106, p. 104492, 2021.

[13] N. Takeda, Z. Li, K. Shige, and O. Terashima, "Study on the machine-learning based system for detecting abnormal pressure drops in hydraulic press machines," The International Journal of Advanced Manufacturing Technology, vol. 130, no. 9, pp. 5045-5054, 2024/02/01 2024, doi: 10.1007/s00170-024-13001-3